\documentclass{article}
\usepackage{amsfonts}
\usepackage{amssymb}
\usepackage{amsmath}
\usepackage{amsthm}
\usepackage{mathtools}
\begin{document}

\title{Conformal Transformations\\ and Cosmological Perturbations\\ in New General Relativity}

\author{Alexey Golovnev$^1$, A. N. Semenova$^2$, V.P. Vandeev$^2$\\
{\small \it $^1$Centre for Theoretical Physics, The British University in Egypt,}\\
{\small \it El Sherouk City, Cairo 11837, Egypt}\\
{\small  agolovnev@yandex.ru}\\
{\small \it $^2$Petersburg Nuclear Physics Institute of National Research Centre ``Kurchatov Institute'',}\\ 
{\small \it Gatchina, 188300, Russia}\\
{\small ala.semenova@gmail.com \hspace{50 pt} vandeev{\_}vp@pnpi.nrcki.ru} } 
\date{}

\maketitle

\begin{abstract}

We study conformal transformations in the most general parity-preserving models of the New General Relativity type. Then we apply them to analysis of cosmological perturbations in the (simplest) spatially flat cosmologies. Strong coupling issues around Minkowski spacetime are seen for many special cases of these models. At the same time, the behaviour of the most general three-parameter case seems to be very robust, presumably always with only the eight first-class constraints coming from diffeomorphisms. Also the case of the so-called 1-parameter New GR doesn't show any discontinuity between Minkowski and the cosmology, though without showing any  deviations from GR which would be observable at this level either.

\end{abstract}

\maketitle

\section{Introduction}

The more we study the modified teleparallel theories, the more serious seem to be their foundational issues. In the highly nonlinear models such as $f(\mathbb T)$, the chaos of remnant symmetries does not allow us to have a universally well-defined number of degrees of freedom \cite{issues}, with the obvious strong coupling of the trivial Minkowski tetrad $\delta^a_{\mu}$ somehow also extending to cosmologies \cite{cosmo1, cosmo2}. Intuitively, one can say that the problem comes from the fact that the Lorentz-symmetry violation of the pure tetrad version comes through a nonlinear function of a boundary term which effectively acts along the gradient of $\mathbb T$ only. This breakdown appears to be not strong enough for giving a new robust number of dynamical modes.

The observation above makes it rather natural to consider other modifications, those which break the symmetries of Teleparallel Equivalent of General Relativity (TEGR) in a more reliable manner. Strangely enough, simple options of taking other quadratic combinations of the torsion tensor components for the Lagrangian density are much less explored than the intricate and problematic nonlinear generalisations. The dynamics of (3-parameter) New General Relativity, or New GR for short, has recently been studied, at the level of Hamiltonian primary constraints \cite{dof1, dof2} and  in the weak gravity limit  \cite{wedof}. In this paper, we would like to add a bit more to this line of research. Namely, we will introduce conformal transformations in New GR and extend our previous work on degrees of freedom around the trivial Minkowski background tetrad \cite{wedof} to the simplest cosmological cases.

To be more precise, the gravity model we study is given by an action
\begin{equation}
S=\frac{1}{16\pi G}\int d^4 x \sqrt{-g} \cdot \mathfrak T
\end{equation}
with the torsion scalar and the superpotential tensor defined as
\begin{equation}
\label{torsionscalar}
{\mathfrak T}=\frac12 T_{\alpha\mu\nu} {\mathfrak S}^{\alpha\mu\nu}, \qquad {\mathfrak S}_{\alpha\mu\nu} = \frac{\mathrm a}{2}\cdot T_{\alpha\mu\nu} +\frac{b}{2}\cdot\left(T_{\mu\alpha\nu} -T_{\nu\alpha\mu} \right)+c\cdot\left( g_{\alpha\mu}T_{\nu} - g_{\alpha\nu}T_{\mu}\right)
\end{equation}
which in the case of ${\mathrm a}=b=c=1$ reproduces the standard $\mathbb T$ scalar of TEGR=GR and $f(\mathbb T)$. We recall that the torsion and contortion tensors are defined via
$$T^{\alpha}_{\hphantom{\alpha}\mu\nu}=\Gamma^{\alpha}_{\mu\nu} - \Gamma^{\alpha}_{\nu\mu}, \qquad T_{\mu}=T^{\alpha}_{\hphantom{\alpha}\mu\alpha}, \qquad K_{\alpha\mu\nu}=\frac12 \left(T_{\alpha\mu\nu}+T_{\mu\alpha\nu}+T_{\nu\alpha\mu}\right)$$
with the connection given as
$$\Gamma^{\alpha}_{\mu\nu}=e^{\alpha}_a \partial_{\mu} e^a_{\nu}$$
in terms of the orthonormal teleparallel tetrad $e^a_{\mu}$.

We will always assume that there is no hypermomentum, and the only source of gravity is a symmetric covariantly-conserved energy-momentum tensor ${\mathcal T}_{\mu\nu}$. Therefore, we can naturally separate the equations \cite{wedof} into antisymmetric and symmetric parts as
\begin{equation}
\label{asymm}
\frac12 {\mathop{\bigtriangledown}\limits^{(0)}}_{\alpha} \left({\mathfrak S}_{\mu\nu}^{\hphantom{\mu\nu}\alpha}-{\mathfrak S}_{\nu\mu}^{\hphantom{\mu\nu}\alpha}\right) - \frac12 \left( K_{\alpha\mu\beta}{\mathfrak S}^{\alpha\hphantom{\nu}\beta}_{\hphantom{\alpha}\nu} -  K_{\alpha\nu\beta}{\mathfrak S}^{\alpha\hphantom{\mu}\beta}_{\hphantom{\alpha}\mu}\right)=0,
\end{equation}
\begin{equation}
\label{symm}
\frac12 {\mathop{\bigtriangledown}\limits^{(0)}}_{\alpha} \left({\mathfrak S}_{\mu}^{\hphantom{\mu}\nu\alpha}+{\mathfrak S}_{\hphantom{\nu}\mu}^{\nu\hphantom{\mu}\alpha}\right) - \frac12 \left( K_{\alpha\mu\beta}{\mathfrak S}^{\alpha\nu\beta} +  K^{\alpha\nu\beta}{\mathfrak S}_{\alpha\mu\beta}\right) + \frac12 {\mathfrak T} \delta^{\nu}_{\mu}=8\pi G \cdot {\mathcal T}^{\nu}_{\mu}
\end{equation}
with ${\mathop{\bigtriangledown}\limits^{(0)}}_{\alpha} $ being the usual Levi-Civita covariant derivative corresponding to the metric
\begin{equation}
\label{mtr}
g_{\mu\nu}=\eta_{ab} e^a_{\mu} e^b_{\nu}.
\end{equation}
Since the left hand side of these equations (\ref{asymm}, \ref{symm}) can be treated as a generalisation of the Einstein tensor, it will be denoted as
\begin{equation}
\label{genEi}
{\mathfrak G}^{\hphantom{\mu}\nu}_{\mu}\equiv  {\mathop\bigtriangledown\limits^{(0)}}_{\alpha} {\mathfrak S}_{\mu}^{\hphantom{\mu}\nu\alpha}- {\mathfrak S}^{\alpha\nu\beta} K_{\alpha\mu\beta} + \frac12 {\mathfrak T}\delta^{\nu}_{\mu}.
\end{equation}
We will show that its conformal transformations can be presented in a relatively compact form.

For the cosmological part of our work, we choose the simplest and most natural spatially-flat teleparallel cosmological background given by, so to say, a conformally trivial tetrad
\begin{equation}
\label{backgr}
e^a_{\mu} = a(\tau)\cdot \delta^a_{\mu}
\end{equation}
which automatically solves the antisymmetric part of equations\footnote{or is compatible with the zero spin-connection, in the "Lorentz-covariant" language \cite{covstuff1, covstuff2, covstuff3}}, respects the cosmological symmetries, and has a conformally flat background metric of 
$$g_{\mu\nu} dx^{\mu} dx^{\nu} = a^2(\tau) \cdot \left( d\tau^2 - dx_i^2 \right).$$
Any change of the time coordinate alone, with no touch of the spatial ones, does leave the matrix of components of the tetrad (\ref{backgr}) diagonal. However, it is very well-known that the use of conformal time is very convenient for the study of cosmological perturbations. And it will also allow us to illustrate the use of conformal transformations in New GR.

In this paper, we will never work in the physical time $dt = a(\tau) d\tau$ and therefore, unlike most of the cosmological literature, we will denote the comformal time derivatives by ${\dot a} \equiv \frac{da}{d\tau}$ (instead of $a^{\prime}$). However, in order to keep the choice of the time variable in mind, the corresponding logarithmic derivative of the scale factor will be denoted by a curly letter: ${\mathcal H}\equiv\frac{\dot a}{a}$.

One can check that the equations of motion with the tetrad Ansatz (\ref{backgr}) are indeed automatically symmetric, and even diagonal, with the only non-trivial components being
\begin{equation}
\label{Friedman}
a^2(\tau)\cdot {\mathfrak G}^0_{\hphantom{0}0} = \frac{3(6c -{\mathrm a} - b)}{4}\cdot  {\mathcal H}^2, \qquad a^2(\tau)\cdot {\mathfrak G}^i_{\hphantom{i}j} = \frac{6c -{\mathrm a} - b}{4}\cdot  (2 {\dot{\mathcal H}} + {\mathcal H}^2) \delta_{ij}.
\end{equation}
Naturally, the case of ${\mathrm a}=b=c=1$ brings us back to GR. More interesting is that, at the background level, all these cosmologies are equivalent to the GR ones, modulo renormalisation of the effective gravitational constant by the factor of $\frac{6c -{\mathrm a} - b}{4}$. 

The case of $6c={\mathrm a}+b$ is very special since this type of cosmology in then possible only in vacuum, though with absolutely any scale factor $a(\tau)$ solving the equations, similar to analogous peculiarity of this model for static spherically symmetric solutions \cite{wespher}. Note that a reason for why the condition of $6c={\mathrm a}+b$ turns out to be so special can be seen by calculating the superpotential trace:
\begin{equation}
\label{supertr}
{\mathfrak S}^{\alpha}_{\hphantom{\alpha}\mu\alpha} = - \frac{6c-{\mathrm a}-b}{2} \cdot T_{\mu}.
\end{equation}
This relation (\ref{supertr}) is very convenient for simplifying many calculations to follow.

In  Section 2 we study the conformal transformations. In Section 3 we introduce the theory of cosmological perturbations and use of conformal transformations for it. In Sections 4, 5, 6 we present the detailed properties of tensor, vector, and scalar perturbations respectively. In Section 7 we conclude.

\section{Conformal transformations in New General Relativity}

Teleparallel theories are no exception to the rule of thumb that, in simple non-linear generalisations of gravity, it is very natural to study the conformal \cite{confo} or even disformal \cite{disfo} transformations. However, the benefit of it is not so obvious for the models like New GR and, to the best of our knowledge, it hasn't been properly investigated yet. Here we will discover that, somewhat surprisingly for us, the conformal transformations make it possible to use a big part of experience with cosmological perturbations in GR for studying those in New GR.

A conformal transformation, and in particular the transformation from weak gravity \cite{wedof} to cosmological physics (\ref{backgr}), can be done via
\begin{equation}
\label{contetr}
e^a_{\mu} \longrightarrow \underline{e^a_{\mu}} = e^{\varphi(x)}\cdot e^a_{\mu},
\end{equation}
with a smooth scalar function $\varphi (x)$. In the Riemannian part, it is the usual and well-known transformation of 
$$g_{\mu\nu}\longrightarrow \underline{g_{\mu\nu}}= e^{2\varphi}\cdot g_{\mu\nu}.$$ 
At the same time, one can easily find that the torsion tensor transforms as
\begin{equation}
\label{contor}
\underline{T^{\alpha}_{\hphantom{\alpha}\mu\nu}} = T^{\alpha}_{\hphantom{\alpha}\mu\nu} + \delta^{\alpha}_{\nu} \partial_{\mu} \varphi - \delta^{\alpha}_{\mu} \partial_{\nu} \varphi, \qquad \underline{T_{\mu}} = T_{\mu} + 3 \partial_{\mu} \varphi.
\end{equation}
An interesting point is that the changes in many other geometric quantities can also be presented in rather nice and simple forms:
\begin{equation}
\label{concont}
\underline{K_{\alpha\mu\nu} }= e^{2\varphi} \cdot \left( K_{\alpha\mu\nu} + g_{\mu\nu} \partial_{\alpha} \varphi - g_{\mu\alpha} \partial_{\nu} \varphi \right), 
\end{equation}
\begin{equation}
\label{consuper}
\underline{{\mathfrak S}_{\alpha\mu\nu} }= e^{2\varphi} \cdot \left( {\mathfrak S}_{\alpha\mu\nu} - \frac{6c-{\mathrm a}-b}{2}\cdot \left( g_{\alpha\nu} \partial_{\mu} \varphi - g_{\alpha\mu} \partial_{\nu} \varphi \right) \right), 
\end{equation}
\begin{equation}
\label{conscal}
\underline{{\mathfrak T} }= e^{-2\varphi} \cdot \left( \vphantom{\frac22} {\mathfrak T} - \frac{6c-{\mathrm a}-b}{2}\cdot \left(2T^{\mu} \partial_{\mu}\varphi+ 3(\partial\varphi)^2 \right)\right)
\end{equation}
where all the quantities in the right hand sides are taken before the transformation.

A little bit less trivial results to check about the transformation (\ref{contetr}) are
\begin{multline}
\label{concovs}
\underline{{\mathop{\bigtriangledown}\limits^{(0)}}_{\alpha}{\mathfrak S}_{\mu}^{\hphantom{\mu}\nu\alpha} }= e^{-2\varphi}\cdot \left({\mathop{\bigtriangledown}\limits^{(0)}}_{\alpha} {\mathfrak S}_{\mu}^{\hphantom{\mu}\nu\alpha} +\left( {\mathfrak S}_{\mu}^{\hphantom{\mu}\nu\alpha} + {\mathfrak S}_{\hphantom{\alpha\nu}\mu}^{\alpha\nu} \right)\partial_{\alpha}\varphi\right.\\
\left. -\frac{6c-{\mathrm a}-b}{2}\cdot \left( {\mathop{\bigtriangledown}\limits^{(0)}}_{\mu}{\mathop{\bigtriangledown^{\nu}}\limits^{(0)}}\varphi-\delta^{\nu}_{\mu} {\mathop{\square}\limits^{(0)}}\varphi - 2(\partial^{\nu}\phi)\partial_{\mu}\varphi  - \delta^{\nu}_{\mu} (\partial\varphi)^2 -T^{\nu}\partial_{\mu}\varphi  \right)\right)
\end{multline}
and 
\begin{multline}
\label{conquad}
\underline{K_{\alpha\mu\beta}{\mathfrak S}^{\alpha\nu\beta} }= e^{-2\varphi}\cdot \left(\vphantom{\frac12} K_{\alpha\mu\beta}{\mathfrak S}^{\alpha\nu\beta} + \left({\mathfrak S}^{\alpha\nu}_{\hphantom{\alpha\nu}\mu} - {\mathfrak S}_{\mu}^{\hphantom{\mu}\nu\alpha}\right)\partial_{\alpha}\varphi\right.\\
 -\left.\frac{6c-{\mathrm a}-b}{2}\cdot \left(\delta^{\nu}_{\mu} (\partial\varphi)^2 - (\partial^{\nu}\varphi)\partial_{\mu}\varphi - K^{\nu\hphantom{\mu}\alpha}_{\hphantom{\nu}\mu}\partial_{\alpha}\varphi\vphantom{\frac12}\right)\right).
\end{multline}
It is slightly cumbersome but absolutely straightforward to do by using the rules of (\ref{contor}, \ref{concont}, \ref{consuper}).

Combining all these expressions (\ref{conscal}, \ref{concovs}, \ref{conquad}), we get
\begin{multline}
\label{congein}
\underline{{\mathfrak G}_{\mu\nu}}= {\mathfrak G}_{\mu\nu} -\frac{6c-{\mathrm a}-b}{2}\cdot \left( {\mathop{\bigtriangledown}\limits^{(0)}}_{\mu}{\mathop{\bigtriangledown_{\nu}}\limits^{(0)}}\varphi-g_{\mu\nu} {\mathop{\square}\limits^{(0)}}\varphi - (\partial_{\mu}\varphi) \partial_{\nu}\varphi -\frac12 g_{\mu\nu}(\partial\varphi)^2 \right)\\
+2{\mathfrak S}_{\mu\nu}^{\hphantom{\mu\nu}\alpha} \partial_{\alpha}\varphi -\frac{6c-{\mathrm a}-b}{2}\cdot \left(K_{\nu\mu}^{\hphantom{\nu\mu}\alpha} \partial_{\alpha}\varphi - T_{\nu}\partial_{\mu}\varphi + g_{\mu\nu} T^{\alpha}\partial_{\alpha}\varphi\right).
\end{multline}
In case of TEGR, ${\mathrm a}=b=c=1$, the first line is the usual conformal transformation of GR 
$$\underline{G_{\mu\nu}}= G_{\mu\nu} -2 {\mathop{\bigtriangledown}\limits^{(0)}}_{\mu}{\mathop{\bigtriangledown_{\nu}}\limits^{(0)}}\varphi +2g_{\mu\nu} {\mathop{\square}\limits^{(0)}}\varphi +2 (\partial_{\mu}\varphi) \partial_{\nu}\varphi + g_{\mu\nu}(\partial\varphi)^2$$
while the second line vanishes due to ${\mathfrak S}_{\mu\nu\alpha}=S_{\mu\nu\alpha}=K_{\nu\mu\alpha}+g_{\mu\nu}T_{\alpha}-g_{\mu\alpha}T_{\nu}$ in this case.

Applying the conformal transformations (\ref{contetr}) has been very useful in $f(\mathbb T)$ theories, for instance for producing the almost "Einstein frame" there. The case of New GR is different since its basic scalars have almost nothing in common with the usual scalar curvature, unlike the standard quantity $\mathbb T$. However, we do see  from the transformation (\ref{congein}) that the conformal rewritings of the gravitational equations (\ref{genEi}) get only a relatively simple contribution on top of what we are used to in GR. It will help us a lot when analysing the cosmological perturbations.

\section{Cosmological perturbations}

In order to get better understanding of the theory and its properties, we are going to analyse the simplest cosmological models (\ref{backgr}) in it assuming the matter content given by a single ideal fluid
\begin{equation}
\label{EMT}
{\mathcal T}_{\mu\nu} = (\rho + p) {\mathfrak v}_{\mu} {\mathfrak v}_{\nu} - p g_{\mu\nu}
\end{equation}
where both the energy density $\rho$ and the pressure $p$ experience fluctuations around the background values. The 4-velocity, ${\mathfrak v}_{\mu} {\mathfrak v}^{\mu}\equiv 1$, is taken as ${\mathfrak v}_{\mu}= a(\tau) \cdot \delta^0_{\mu}$ at the background level, with the perturbation
$${\mathfrak v}_i = \partial_i {\mathfrak v} + {\tilde{\mathfrak v}}_i, \qquad \partial_i {\tilde{\mathfrak v}}_i=0$$
decomposed into the scalar and vector parts, i.e. potential and vortical flows. 

The linear perturbation quantities in the right hand side of the cosmological equations will be
\begin{equation}
\label{emp}
\delta {\mathcal T}^0_0 = \delta\rho, \qquad \delta {\mathcal T}^0_i = (\rho + p)\cdot \frac{{\mathfrak v}_i}{a(\tau)}, \qquad \delta {\mathcal T}^i_j = - \delta p\cdot \delta_{ij},
\end{equation}
and we will always assume that those satisfy the energy-momentum conservation laws which is anyway required by a diffeomorphism-invariant gravity theory, see below the sections on vector and scalar perturbations for a bit more details.

\subsection{Generalities of the perturbative calculations}

We will nor care about the position of indices in the perturbative notations (see the $\delta_{ij}$ symbol above), for we will keep track of only the linear spatial (Euclidean) covariance. Therefore, from the perturbation of the co-tetrad represented as
\begin{equation}
\label{tetpert}
e^a_{\mu}= a(\tau) \cdot \left(I + {\mathfrak e}^a_{\mu}\right)
\end{equation}
with $I$ being the unit matrix, we deduce the tetrad perturbation to be
$$e^{\mu}_a = \frac{1}{a(\tau)}\cdot \left(I - {\mathfrak e}^a_{\mu}\right) + {\mathcal O}({\mathfrak e}^2).$$

In principle, this is all what we need. The torsion tensor components then follow as
\begin{equation}
\label{torpert}
\begin{array}{rcl}
T_{00i}= -T_{0i0} & = &  a^2 \cdot \left({\dot{\mathfrak e}}^0_i - \partial_i {\mathfrak e}^0_0 + {\mathcal H} \left({\mathfrak e}^0_i-{\mathfrak e}^i_0\right)\right),\\
T_{0ij} & = & a^2 \cdot \left( \partial_i {\mathfrak e}^0_j - \partial_j {\mathfrak e}^0_i\right),\\
T_{ijk} & = & a^2 \cdot \left( \partial_k {\mathfrak e}^i_j - \partial_j {\mathfrak e}^i_k\right),\\
T_{i0j} =- T_{ij0}& = & a^2 \cdot \left( - {\mathcal H} \delta_{ij}- {\dot{\mathfrak e}}^i_j + \partial_j {\mathfrak e}^i_0 -  {\mathcal H} \left({\mathfrak e}^i_j + {\mathfrak e}^j_i\right)\right),
\end{array}
\end{equation}
to the linear order in perturbations. Obviously, if one wants to have a look at the Minkowski spacetime, it is the simple limit of $\mathcal H =0$ which we have treated in the previous paper \cite{wedof}.

At the same time, using the linear approximation of the metric (\ref{mtr})
\begin{equation}
\label{metpert}
g_{00}= a^2 \cdot \left(1 + 2 {\mathfrak e}^0_0\right), \qquad g_{0i}= a^2 \cdot \left({\mathfrak e}^0_i - {\mathfrak e}^i_0\right), \qquad g_{ij}= - a^2 \cdot \left(\delta_{ij} + {\mathfrak e}^i_j + {\mathfrak e}^j_i\right)
\end{equation}
and the inverse metric
\begin{equation}
\label{invmetr}
g^{00}= \frac{1}{a^2} \cdot \left(1 - 2 {\mathfrak e}^0_0\right), \qquad g^{0i}= \frac{1}{a^2} \cdot \left({\mathfrak e}^0_i - {\mathfrak e}^i_0\right), \qquad g^{ij}= - \frac{1}{a^2} \cdot \left(\delta_{ij} - {\mathfrak e}^i_j - {\mathfrak e}^j_i\right)
\end{equation}
corresponding to the tetrad (\ref{tetpert}), one can raise or lower any indices of the tensors; and all the rest of teleparallel quantities are simple algebraic combinations of the torsion tensor components above. For example, we can find the torsion vector:
\begin{equation}
\label{torvec}
T_0= 3{\mathcal H}  +{\dot{\mathfrak e}}^i_i - \partial_i {\mathfrak e}^i_0, \qquad T_i= - {\dot{\mathfrak e}}^0_i + \partial_i {\mathfrak e}^0_0 - \partial_j {\mathfrak e}^j_i + \partial_i {\mathfrak e}^j_j,
\end{equation}
summation over the repeated spatial indices is assumed. For the Levi-Civita covariant derivatives, one would also need to calculate the corresponding connection coefficients for the metric (\ref{metpert}) but we will not explicitly write them here for the procedure is pretty standard and simple.

\subsection{Parametrisation}

As usual, one can separate the perturbations into scalar, vector, and tensor parts. They won't mix with each other at the linear level. The usual way to parametrise them \cite{cosmo1, wedof} is as follows:
\begin{equation}
\label{tetrpert-old}
\begin{array}{rcl}
{\mathfrak e}^0_0 & = & \phi\\
{\mathfrak e}^0_i & = & \partial_i \beta+u_i\\
{\mathfrak e}^i_0 & = & \partial_i \zeta+v_i\\
{\mathfrak e}^i_j & = & -\psi \delta_{ij}+\partial^2_{ij}\sigma+\epsilon_{ijk}\partial_k s+\partial_j c_i+\epsilon_{ijk}w_k+\frac12 h_{ij}.
\end{array}
\end{equation}
The newly introduced variables are six scalars $\phi,\beta,\zeta,\psi,\sigma,s$, four divergenceless ($\partial_i {\mathbb V}_i =0$) vectors $u_i,v_i,c_i,w_k$, and one tensor $h_{ij}$ which is symmetric $h_{ij}=h_{ji}$, traceless $h_{ii}=0$, and totally divergenceless $\partial_{i} h_{ij}=0$. 

For the sake of reader's convenience, we will show all the torsion tensor components in the standard parametrisation (\ref{tetrpert-old}). However, following our previous paper \cite{wedof}, we also introduce the convenient reparametrisation of
$$w_k=\epsilon_{ijk} \partial_i \chi_j, \qquad {\mathcal M}_i=\frac{u_i -v_i}{2}, \qquad {\mathcal L}_i=\frac{u_i + v_i}{2}.$$
Altogether it means
\begin{equation}
\label{tetrpert}
\begin{array}{rcl}
{\mathfrak e}^0_0 & = & \phi\\
{\mathfrak e}^0_i & = & \partial_i \beta +{\mathcal L}_i +{\mathcal M}_i\\
{\mathfrak e}^i_0 & = & \partial_i \zeta+{\mathcal L}_i -{\mathcal M}_i\\
{\mathfrak e}^i_j & = & -\psi \delta_{ij}+\partial^2_{ij}\sigma+\epsilon_{ijk}\partial_k s+\partial_j c_i+\partial_i\chi_j - \partial_j\chi_i+\frac12 h_{ij}.
\end{array}
\end{equation}

Note that on top of the usual metric perturbations ($\phi$, $\psi$, $\beta-\zeta$, $\sigma$, $u_i - v_i = 2{\mathcal M}_i$, $c_i$, $h_{ij}$), we have an arbitrary Lorentz transformation of the orthonormal tetrad (\ref{tetpert}): rotations around 
$$\partial_k s + w_k = \partial_k s +  \epsilon_{ijk}\partial_i\chi_j$$ 
and boosts along 
$$\partial_i (\beta + \zeta) + u_i + v_i = \partial_i (\beta + \zeta) +2{\mathcal L}_i.$$

\subsubsection{Gauge symmetry}

In all the models under consideration there is the gauge symmetry \cite{cosmo1, wedof} of infinitesimal coordinate transformations, $x^{\mu}\longrightarrow x^{\mu} + \xi^{\mu}(x)$. In relation to the spatial rotations, the temporal component $\xi^0$ is a scalar, while the spatial part can be decomposed into a scalar and a (divergenceless) vector as $\xi^i=\partial_i\xi + {\tilde \xi}_i$, too. If the co-tetrad is transformed as a collection of 1-forms $e^a$, then it takes the form of \cite{cosmo1}
\begin{equation}
\label{gaugetrans}
\begin{array}{rcl}
\phi & \longrightarrow & \phi-{\dot \xi}^0-{\mathcal H}\xi^0\\
\psi & \longrightarrow & \psi+ {\mathcal H}\xi^0\\
\sigma & \longrightarrow & \sigma-\xi\\
\beta & \longrightarrow & \beta-\xi^0\\
\zeta & \longrightarrow & \zeta-\dot\xi\\
c_i & \longrightarrow & c_i -{\tilde\xi}_i\\
v_i & \longrightarrow & v_i -{\dot {\tilde\xi}_i}.
\end{array}
\end{equation}

The simplest choice of gauge invariant variables is
\begin{equation}
\label{gaugeinv}
s, \quad Z=\zeta - \dot\sigma, \quad \Phi=\phi - \dot\beta - {\mathcal H}\beta, \quad \Psi=\psi + {\mathcal H} \beta, \quad w_i, \quad u_i, \quad V_i=v_i - {\dot c}_i, \quad h_{ij}
\end{equation}
which are four scalars, three vectors and one tensor summing together to $4 + 2\times 3 + 2 = 12$ variables, precisely as it must be after subtracting $4$ gauge freedoms from $16$ components. Note that the usual six Bardeen variables can be found in terms of them (\ref{gaugeinv}) as
\begin{equation}
\label{Bardeen}
\Phi_B = \Phi + \dot Z + {\mathcal H} Z, \qquad \Psi_B = \Psi - {\mathcal H} Z, \qquad  {V_B}_i = V_i - u_i,\qquad  h_{ij}
\end{equation}
with six Lorentz parameters on top.

Our gauge choice \cite{cosmo1, wedof} will be
\begin{equation}
\label{ourgauge}
\sigma=0, \qquad \beta=\zeta, \qquad c_i=0
\end{equation}
which is a natural generalisation of the conformal Newtonian gauge. Precisely as the latter, it is very nice for not leaving any remnant symmetries. Its non-invariant quantities $\phi, \psi, \zeta, v_i$ are equal (in this particular gauge) to the gauge-invariant variables, $\Phi_B, \Psi_B, Z, V_i$ respectively (\ref{gaugeinv}, \ref{Bardeen}).

\subsection{The modes around Minkowski}

The count of degrees of freedom was performed around the Minkowski spacetime in our previous paper \cite{wedof}, in the chosen diffeo-gauge (\ref{ourgauge}). It is appropriate to keep it in mind when analysing the cosmological perturbations in order to detect (some of) the strong coupling issues. In principle, all those results are just a special case of $\mathcal H=0$ (and no matter) of what will follow in the next sections. Nevertheless, we give here a summary of a possible dynamical classification of the variables (\ref{tetrpert}) which are left after gauge fixing (\ref{ourgauge}).

\begin{center}
{\bf Degrees of freedom in Minkowski vacuum} 
\begin{tabular}{c|c|c|c|c|} 
 \hline
& parameter range & dynamical & constrained &  still pure gauge \\ 
\hline
Type 1 & generic & $h_{ij}$, ${\mathcal M}_i$, $\chi_i$, $\zeta  $, $s$ & ${\mathcal L}_i$, $\phi$, $\psi$ & ---\\ 
\hline
Type 2 & ${\mathrm a}+b =2c $ & $h_{ij}$, $s$ & ${\mathcal M}_i$, ${\mathcal L}_i$, $\phi$, $\psi$ & $\chi_i$, $\zeta$\\
\hline
Type 3 & ${\mathrm a}=b$ & $h_{ij}$, $\zeta$ & ${\mathcal M}_i$, $\chi_i$, $\phi$, $\psi$ & ${\mathcal L}_i$, $s$\\
\hline
Type 4 & ${\mathrm a}+b =0$ & half of ${\mathcal M}_i$, $s$ & half of ${\mathcal M}_i$, ${\mathcal L}_i$, $\phi$, $\zeta$ & $h_{ij}$, $\chi_i$, $\psi$ \\
\hline
Type 5 & ${\mathrm a}+b =6c $ & $h_{ij}$, ${\mathcal M}_i$,  $\chi_i$, $s$ & ${\mathcal L}_i$, $\phi+\psi$, $\zeta$ & $\phi-\psi$\\
\hline
Type 6 & ${\mathrm a}=b=c$ & $h_{ij}$ & ${\mathcal M}_i$, $\phi$, $\psi$  & ${\mathcal L}_i$, $\chi_i$, $\zeta  $, $s$ \\
\hline
Type 7 & ${\mathrm a}=b=3c$ & $h_{ij}$ & ${\mathcal M}_i$,  ${\chi_i}$, $\phi+\psi$,  $\zeta$ & ${\mathcal L}_i$, $\phi-\psi$, $s$\\
\hline
Type 8 & ${\mathrm a}=b=0$ & --- & $\chi_i$, $\phi$, $\zeta$ & $h_{ij}$, ${\mathcal M}_i$, ${\mathcal L}_i$, $\psi$,   $s$\\
\hline
Type 9 & ${\mathrm a}+b =c=0$ & $s$ &  ${\mathcal L}_i$ &  $h_{ij}$, ${\mathcal M}_i$, $\chi_i$, $\phi$, $\psi$, $\zeta$\\
\hline
\end{tabular}
\end{center}

Note that this classification should not be taken too literally. For example, when a combination of $\dot\chi_i + {\mathcal L}_i$ was fully constrained with no more information on the involved variables, we interpreted it as $\chi_i$ being pure gauge and ${\mathcal L}_i$ constrained. In this case, the choice was dictated by avoidance of a fake Cauchy datum, though it was often pretty voluntaristic, too. The separation of ${\mathcal M}_i$ into two halves in one of the cases meant having  a constraint on $\dot{\mathcal M}_i$, leaving one half of the necessary Cauchy data for ${\mathcal M}_i$.

\subsection{The use of conformal transformations}

The brute force approach to cosmological perturbations is definitely possible though quite cumbersome. However, we can take the rather simple weak gravity regime \cite{wedof}, i.e. $a(\tau)=1$ and $\mathcal H =0$, whose linearised equations, being around the trivial tetrad 
$$e^a_{\mu}=\delta^a_{\mu} + {\mathfrak e}^a_{\mu},$$
get reduced to partial derivatives of the superpotential,
$$ {\mathfrak G}^{\hphantom{\mu}\nu}_{\mu} =  {\mathop\bigtriangledown\limits^{(0)}}_{\alpha} {\mathfrak S}_{\mu}^{\hphantom{\mu}\nu\alpha}- {\mathfrak S}^{\alpha\nu\beta} K_{\alpha\mu\beta} + \frac12 {\mathfrak T}\delta^{\nu}_{\mu} = \partial_{\alpha} {\mathfrak S}_{\mu}^{\hphantom{\mu}\nu\alpha} + {\mathcal O}(\mathfrak e^2) $$
with the positions of all the indices manipulated by the simple $\eta_{\mu\nu}$ metric, and then we apply the conformal transformation (\ref{congein}) to it. Note that our conformal factor of $a^2(\tau)$ has got the property of 
$$\partial_{\mu} \varphi = {\mathcal H}\cdot \delta^0_{\mu}.$$

Anyone who has ever approached cosmological perturbations in GR via conformal transformations can immediately recall that
$$(\partial\varphi)^2 = {\mathcal H}^2 \cdot g^{00}, \qquad {\mathop{\bigtriangledown}\limits^{(0)}}_{\mu}{\mathop{\bigtriangledown_{\nu}}\limits^{(0)}}\varphi = \dot{\mathcal H}\cdot \delta^0_{\mu}\delta^0_{\nu} - {\mathcal H}\cdot {\mathop{\Gamma}\limits^{(0)}}{}^0_{\mu\nu}, \qquad {\mathop{\square}\limits^{(0)}}\varphi =\dot{\mathcal H}\cdot g^{00} - {\mathcal H}\cdot g^{\alpha\beta} {\mathop{\Gamma}\limits^{(0)}}{}^0_{\alpha\beta},$$
the formulae which are anyway quite obvious. The Levi-Civita connection coefficients correspond to the perturbed Minkowski metric and are quite easy to calculate.

Therefore, we get the final version of the cosmological gravity tensor (\ref{congein}) as
\begin{multline}
\label{cosmgein}
\underline{{\mathfrak G}_{\mu\nu}} = {\mathfrak G}_{\mu\nu} -\frac{6c-{\mathrm a}-b}{2}\cdot\left( \dot{\mathcal H}\cdot \left( \delta^0_{\mu} \delta^0_{\nu} - g^{00}g_{\mu\nu}\right) - {\mathcal H}^2\cdot \left( \delta^0_{\mu} \delta^0_{\nu} +\frac12 g^{00}g_{\mu\nu}\right) \right)\\
+ {\mathcal H}\cdot \left(2 {\mathfrak S}_{\mu\nu}^{\hphantom{\mu\nu}0} -\frac{6c -{\mathrm a}-b}{2}\cdot \left(K_{\nu\mu}^{\hphantom{\nu\mu}0} - T_{\nu}\delta^0_{\mu}+ g_{\mu\nu} T^0 +g_{\mu\nu} g^{\alpha\beta} {\mathop{\Gamma}\limits^{(0)}}{}^0_{\alpha\beta} -{\mathop{\Gamma}\limits^{(0)}}{}^0_{\mu\nu}\right)\right).
\end{multline}
The nice property of going from weak gravity to cosmological perturbations like this is  that all the quantities in the right hand side, except the metric, are zero at the background level, therefore the first order approximation is very elementary and straightforward.

Let us separate the relation (\ref{cosmgein}) into symmetric and antisymmetric parts. In the symmetric part (\ref{symm}) of equations
$$\underline{{\mathfrak G}_{\mu\nu}} =8\pi G\cdot {\mathcal T}_{\mu\nu},$$ 
it makes sense to raise one of the indices, due to the nice shape of an ideal fluid energy-momentum tensor (\ref{emp}) and also for some calculations becoming more compact,
\begin{multline}
\label{scosmgein}
\frac{ {\mathfrak G}^{\mu}_{\hphantom{\mu}\nu} + {\mathfrak G}^{\hphantom{\mu}\mu}_{\nu}}{2} -\frac{6c-{\mathrm a}-b}{2}\cdot\left( \dot{\mathcal H}\cdot \left( g^{\mu 0} \delta^0_{\nu} - g^{00}\delta^{\mu}_{\nu}\right) - {\mathcal H}^2\cdot \left( g^{\mu 0} \delta^0_{\nu} +\frac12 g^{00}\delta^{\mu}_{\nu}\right) +  {\mathcal H}\cdot \left(\delta^{\mu}_{\nu} g^{\alpha\beta} {\mathop{\Gamma}\limits^{(0)}}{}^0_{\alpha\beta} -g^{\mu\alpha}{\mathop{\Gamma}\limits^{(0)}}{}^0_{\alpha\nu}\right)\right)\\
+ {\mathcal H}\cdot \left( \frac{3({\mathrm a}+b-2c)}{4}\cdot \left(T_{\hphantom{\mu}\nu}^{\mu\hphantom{\nu}0} +T_{\nu}^{\hphantom{\nu}\mu 0}\right) +\frac{{\mathrm a}+b-2c}{4}\cdot \left(2\delta^{\mu}_{\nu}T^0 - g^{\mu 0} T_{\nu} - \delta^0_{\nu} T^{\mu}\right)\right) = 8\pi G a^2(\tau) \cdot {\mathcal T}^{\mu}_{\nu}.
\end{multline}
The first line in it (\ref{scosmgein}) basically repeats what we have got in GR, while the only genuinely new contribution is in the second line. At the same time, the antisymmetric part (\ref{asymm}) can easily be transformed into
\begin{equation}
\label{ascosmgein}
\frac{{\mathfrak G}_{\mu\nu}-{\mathfrak G}_{\nu\mu}}{2} + {\mathcal H}\cdot \left( \frac{{\mathrm a}-b}{2}\cdot \left(T_{\mu\nu}^{\hphantom{\mu\nu}0} -T_{\nu\mu}^{\hphantom{\nu\mu}0}\right) +\frac{6c-{\mathrm a}-5b}{4}\cdot T^0_{\hphantom{0}\mu\nu} + \frac{{\mathrm a}+b-2c}{4}\cdot \left(T_{\mu}\delta^0_{\nu}- T_{\nu}\delta^0_{\mu} \right)\right) =0.
\end{equation}
We will apply these results (\ref{scosmgein}, \ref{ascosmgein}) to all the three perturbation sectors separately.

\section{The tensor sector}

As always, the tensor sector is the simplest one, and it can easily be done even directly around the cosmological spacetime. The only non-vanishing basic quantities are the torsion tensor components \cite{cosmo1}
\begin{equation}
\begin{array}{rcl}
T_{ijk} & =  & \frac{a^2}{2}\cdot \left(\partial_k h_{ij} - \partial_j h_{ik}\right),\\
-T_{i0j} = T_{ij0} & = & a^2\cdot {\mathcal H} \delta_{ij} + \frac{a^2}{2}\cdot \left({\dot h}_{ij}+ 2{\mathcal H}h_{ij}\right),
\end{array}
\end{equation}
and the metric
$$a^2(\tau) \cdot \left(d\tau^2 - (\delta_{ij} + h_{ij}) dx^i dx^j\right).$$
This sector is gauge invariant and can contribute to the symmetric spatial equation only. If the matter content is an ideal fluid, then the usual GWs have no source term:
\begin{equation}
\label{GWeq}
a^2(\tau)\cdot \delta {\mathfrak G}^i_{\hphantom{i}j }= - \frac{{\mathrm a}+b}{4}\cdot \left({\ddot h}_{ij} + 2{\mathcal H} {\dot h}_{ij} -\bigtriangleup h_{ij}\right)=0.
\end{equation}
In both Minkowski and cosmological cases, there are two tensor polarisations which are dynamical as long as ${\mathrm a}+b\neq 0$, and otherwise become pure gauge in the linearised theory.

\subsection{Approach from the case of Minkowski}

For other sectors, it will be very helpful to go another way. The weak gravity limit is that of $a(\tau)\equiv 1$. One can immediately see that the only non-trivial components of the tensorial equations are given by
$${\mathfrak G}^i_{\hphantom{i}j } = - \frac{{\mathrm a}+b}{4}\cdot \left({\ddot h}_{ij}  -\bigtriangleup h_{ij}\right) $$
and are always symmetric. In the conformal transformation to cosmology (\ref{scosmgein}), we find that
$${\mathcal H}\cdot \left(\frac{{\mathrm a}+b-6c}{2}\cdot {\mathop{\Gamma}\limits^{(0)}}{}^0_{ij} - \frac{3({\mathrm a}+b-2c)}{4}\cdot \left(T_{ij0}+T_{ji0}\right)  \right)= -\frac{{\mathrm a}+b}{2}\cdot {\mathcal H} {\dot h}_{ij}$$
which yields the final result of
$$ a^2(\tau)\cdot \underline{{\mathfrak G}^i_{\hphantom{i}j}} = \frac{6c -{\mathrm a} - b}{4}\cdot  (2 {\dot{\mathcal H}} + {\mathcal H}^2) \delta_{ij}- \frac{{\mathrm a}+b}{4}\cdot \left({\ddot h}_{ij} + 2{\mathcal H} {\dot h}_{ij} -\bigtriangleup h_{ij}\right)$$
reproducing the equation (\ref{GWeq}).

\section{The vector sector}

In the vector sector of parametrisation (\ref{tetrpert-old}), the torsion tensor components \cite{cosmo1} take the form of
\begin{equation}
\label{vecttor}
\begin{array}{rcl}
T_{00i} & = & a^2 \cdot \left({\dot u}_i + {\mathcal H}(u_i - v_i) \right),\\
T_{0ij} & = &  a^2 \cdot \left(\partial_i u_j - \partial_j u_i\right),\\
T_{ijk} & = &  a^2 \cdot \left(\epsilon_{ijl}\partial_k w_l - \epsilon_{ikl}\partial_j w_l \right),\\
T_{i0j} & = & -a^2 \cdot {\mathcal H} \delta_{ij} +a^2 \cdot \left(\partial_j v_i  - \partial_j {\dot c}_i - {\mathcal H} (\partial_j c_i + \partial_j c_i) - \epsilon_{ijk}{\dot w}_k\right), \\
{} & {} & \xrightarrow{\text{gauge fixing}} -a^2\cdot {\mathcal H} \delta_{ij} +a^2 \cdot \left(\partial_j v_i - \epsilon_{ijk}{\dot w}_k \right)
\end{array}
\end{equation}
with the torsion vector being equal to
\begin{equation}
\label{vectvec}
T_0= 3{\mathcal H}, \qquad T_i = - {\dot u}_i + \epsilon_{ijk} \partial_j w_k.
\end{equation}

Even though the metric in our gauge (\ref{ourgauge}), $c_i=0$,
$$a^2(\tau)\cdot \left(d\tau^2 + 2 (u_i - v_i )d\tau dx^i - dx^i dx^i\right) =a^2(\tau)\cdot \left(d\tau^2 + 4 {\mathcal M}_i d\tau dx^i - dx^i dx^i\right) $$
with $g_{0i}=  a^2(\tau) \cdot 2{\mathcal M}_i$, looks quite simple, its Levi-Civita connection coefficients are non-trivial and it is easier to approach the perturbation equations via the conformal transformation (\ref{cosmgein}) from the weak gravity limit. On the other hand, the final expressions are relatively compact, as compared to the scalar case, and therefore we will give a bit more of explicit details when analysing the degrees of freedom in the vector case, for illustrative purposes.

\subsection{The weak gravity case}

To make the derivations easier, we indeed start from Minkowski space which is $a(\tau)\equiv 1$. Taking the ${\mathcal H}=0$ limit of the formulae (\ref{vecttor}), and substituting $w_i$ by $\chi_i$ and $u_i$ and $v_i$ by ${\mathcal M}_i$ and ${\mathcal L}_i$, as in the formulae (\ref{tetrpert}), into the torsion tensor
\begin{equation}
\label{vecttorf}
\begin{array}{rcl}
-T_{0i0}=T_{00i} & = & \dot{\mathcal L}_i + \dot{\mathcal M}_i,\\
T_{0ij} & = &  \partial_i \left( {\mathcal L}_j + {\mathcal M}_j \right)- \partial_j \left( {\mathcal L}_i + {\mathcal M}_i \right),\\
T_{ijk} & = & \partial_i \left(\partial_k \chi_j - \partial_j  \chi_k \right),\\
-T_{ij0}=T_{i0j} & = &  \partial_j\left( {\mathcal L}_i - {\mathcal M}_i + \dot\chi_i \right) - \partial_i \dot\chi_j,
\end{array}
\end{equation}
and the torsion vector
$$T_0 =0, \qquad T_i = - \dot{\mathcal L}_i - \dot{\mathcal M}_i - \bigtriangleup \chi_i,$$
we calculate the ${\mathfrak G}_{\mu\nu}\approx \partial_{\alpha} {\mathfrak S}_{\mu\nu}^{\hphantom{\mu\nu}\alpha}$ components:
$${\mathfrak G}_{0i} = \frac{2c-{\mathrm a}-b}{2}\cdot \left(\ddot{\mathcal L}_i +  \ddot{\mathcal M}_i\right) + \bigtriangleup \left(\frac{{\mathrm a}-b}{2} \cdot {\mathcal L}_i + \frac{{\mathrm a}+b}{2} \cdot {\mathcal M}_i + (c-b)\cdot \dot\chi_i\right),$$
$${\mathfrak G}_{i0} =  \bigtriangleup \left(\frac{b-{\mathrm a}}{2} \cdot {\mathcal L}_i + \frac{{\mathrm a}+b}{2} \cdot {\mathcal M}_i + \frac{b-{\mathrm a}}{2}\cdot \dot\chi_i\right),$$
$${\mathfrak G}_{ij} =  \frac{{\mathrm a}-b}{2}\cdot \left(\partial_i\ddot\chi_j-\partial_j\ddot\chi_i\right) +\frac{2c-{\mathrm a}-b}{2}\cdot\partial_i\bigtriangleup\chi_j + c\cdot \partial_i \dot{\mathcal M}_j + \frac{{\mathrm a}+b}{2}\cdot \partial_j \dot{\mathcal M}_i + (c-b)\cdot \partial_i \dot{\mathcal L}_j + \frac{b-{\mathrm a}}{2}\cdot \partial_j \dot{\mathcal L}_i$$
to be used in the equations of motion. The expected redundancy of these equations is immediately obvious, for $\partial_j {\mathfrak G}_{ij} \equiv \dot{\mathfrak G}_{i0}$ due to antisymmetry of ${\mathfrak S}_{\alpha\mu\nu}$.

The difference from our previous paper \cite{wedof} is twofold. First, as is already mentioned above, we work in the parametrisation (\ref{tetrpert}) with $\chi_i$, $\mathcal L_i$ and $\mathcal M_i$ right away. In particular, it is immediately obvious that all the Lorentzian variables ($\mathcal L_i$ and $\chi_i$) disappear from the equations when ${\mathrm a}=b=c$. Second, for the reader's convenience, we have presented the tensor components themselves instead of their symmetric and antisymmetric parts which can be seen directly from these formulae, of course. The symmetry of the tensor ${\mathfrak G}_{\mu\nu}$ at the locus of ${\mathrm a}=b=c$ is anyway evident.

\subsection{Conformal transformation to cosmology}

The conformal transformation (\ref{cosmgein}) is not difficult to implement. For example, the mixed antisymmetric components $\frac12 \left({\mathfrak G}_{0i} - {\mathfrak G}_{i0}\right)$ get an addition of $\frac{2c-{\mathrm a}-b}{4}\cdot{\mathcal H} (3T_{00i}+T_i)$ resulting in the equation
\begin{equation}
\label{mixasym}
(2c-{\mathrm a}-b)\cdot \left(\ddot{\mathcal L}_i + \ddot{\mathcal M}_i +2{\mathcal H} \left(\dot{\mathcal L}_i + \dot{\mathcal M}_i \right) - {\mathcal H}\bigtriangleup\chi_i\right) +\bigtriangleup \left(\vphantom{\dot{\mathcal L}_i}  2 ({\mathrm a}-b) \cdot {\mathcal L}_i +(2c-3b+{\mathrm a})\cdot\dot\chi_i \right)=0
\end{equation}
upon multiplication by four.

Analogously, the spatial antisymmetric components $\frac12 \left(\underline{{\mathfrak G}_{ij}} - \underline{{\mathfrak G}_{ji}}\right)=0$ take the form of $\partial_i{\mathbb V}_j - \partial_j{\mathbb V}_i=0 $ which implies $\bigtriangleup {\mathbb V}_i=0$. We solve it as $\mathbb V_i=0$ which gives
\begin{equation}
\label{spatasym}
2({\mathrm a}-b)\cdot \left(\ddot\chi_i + 2{\mathcal H}\dot\chi_i\right) + (2c-{\mathrm a}-b)\cdot \left(\bigtriangleup \chi_i + \dot{\mathcal M}_i +3{\mathcal H}{\mathcal M}_i\right)+ (2c-3b+{\mathrm a})\cdot \dot{\mathcal L}_i + (6c-7b+{\mathrm a})\cdot {\mathcal H}{\mathcal L}_i=0
\end{equation}
with both antisymmetric equations (\ref{mixasym}, {\ref{spatasym}}) still vanishing at ${\mathrm a}=b=c$.

In the symmetric part of mixed components, we also get a contribution of $\frac{2c-{\mathrm a}-b}{4}\cdot{\mathcal H} (3T_{00i}+T_i)$ which means that it is only the  ${\mathfrak G}_{0i}$ component, and not the ${\mathfrak G}_{i0}$ one, which gets transformed. Upon multiplication by four, as in all the other components, we get
\begin{equation}
\label{mixsym}
(2c-{\mathrm a}-b)\cdot \left(\ddot{\mathcal L}_i + \ddot{\mathcal M}_i +2{\mathcal H} \left(\dot{\mathcal L}_i + \dot{\mathcal M}_i \right) +\bigtriangleup \left(\dot\chi_i - {\mathcal H}\chi_i\right)\right) + 2({\mathrm a}+b)\cdot \bigtriangleup {\mathcal M}_i = 32\pi G a(\tau) \cdot (\rho + p)\tilde{\mathfrak v}_i
\end{equation}
with $\frac{1}{a(\tau)}(\rho + p)\tilde{\mathfrak v}_i$ representing the divergenceless part of the mixed energy-momentum tensor components (\ref{emp}).

In the symmetric spatial part, the novel contribution of New GR is $\frac{3(2c-{\mathrm a}-b)}{4}\cdot {\mathcal H} (T_{i0j}+T_{j0i})$, and therefore we get
$$a^2(\tau) \cdot\frac12 \left(\underline{{\mathfrak G}^i {}_j}+\underline{{\mathfrak G}_j {}^i}\right) = \frac{6c-{\mathrm a}-b}{4}(2\dot{\mathcal H} + {\mathcal H}^2)\delta_{ij} - (\partial_i {\mathbb V}_j + \partial_j {\mathbb V}_i)$$
$$\mathrm{with}\quad {\mathbb V}_i = \frac{2c-{\mathrm a}-b}{4}\cdot \left(\dot{\mathcal L}_i + 3{\mathcal H}{\mathcal L}_i +\bigtriangleup \chi_i\right) + \frac{2c+{\mathrm a}+b}{4}\cdot \dot{\mathcal M}_i + \frac{6c+{\mathrm a}+b}{4}\cdot {\mathcal H}{\mathcal M}_i$$
by also using ${\mathop{\Gamma}\limits^{(0)}}{}^0_{ij} =\partial_i {\mathcal M}_j + \partial_j {\mathcal M}_i$. As long as the matter content can be taken as an ideal fluid, the source term vanishes and we solve it as
\begin{equation}
\label{spatsym}
(2c-{\mathrm a}-b)\cdot \left(\dot{\mathcal L}_i + 3{\mathcal H}{\mathcal L}_i +\bigtriangleup \chi_i\right) + (2c+{\mathrm a}+b)\cdot \dot{\mathcal M}_i  + (6c+{\mathrm a}+b)\cdot {\mathcal H}{\mathcal M}_i=0
\end{equation}
thus finalising the whole system of equations (\ref{mixasym}, \ref{spatasym}, \ref{mixsym}, \ref{spatsym}).

\subsection{Analysis of equations}

As always in a gauge-fixed system, we encounter more equations (\ref{mixasym}, \ref{spatasym}, \ref{mixsym}, \ref{spatsym}) than variables (${\mathcal M}_i, {\mathcal L}_i, \chi_i$). One can easily find the dependence among the equations by checking that the left hand sides satisfy an identical relation 
\begin{equation}
\label{vecBia}
(\partial_0  +  2{\mathcal H}) \left( \boxed{\ref{mixsym}} - \boxed{\ref{mixasym}} \right) - \left(\boxed{\ref{spatsym}} - \boxed{\ref{spatasym}} \right) \equiv 0
\end{equation}
where $\boxed{N}$ denotes the gravitational side of the equation number $N$. This is a direct consequence of the spatial components of generalised Bianchi identity \cite{wedof, HaSh, meBia}
\begin{equation}
\label{Bianchi}
{\mathop{\bigtriangledown}\limits^{(0)}}_{\mu} {\mathfrak G}^{\alpha\mu} + K^{\mu\alpha\nu} {\mathfrak G}_{\mu\nu} \equiv 0.
\end{equation}
This redundancy (\ref{vecBia}) of equations means that one of them can safely be neglected, as long as we impose the "conservation" ${\mathop{\bigtriangledown}\limits^{(0)}}_{\alpha} {\mathcal T}^{\alpha}_i = 0$ on the energy-momentum tensor, or in other words
\begin{equation}
\label{veccons}
\frac{\tilde{\mathfrak v}_i}{a(\tau)} \propto \frac{1}{a^4({\tau})\cdot (\rho +p) }
\end{equation}
precisely as in GR implying for the right hand side of the equation (\ref{mixsym}):
$$\partial_0 \left( \vphantom{\int} a(\tau) \cdot (\rho + p)\tilde{\mathfrak v}_i\right) = -2 {\mathcal H} a(\tau) \cdot (\rho + p)\tilde{\mathfrak v}_i.$$ 
This decay of vortical motions during expansion is nothing but  a reflection of the flat-space angular momentum conservation. If it was not the case, it would come into contradiction with the identical vanishing of the combination (\ref{vecBia}) of the left hand sides of equations (\ref{mixasym}, \ref{spatasym}, \ref{mixsym}, \ref{spatsym}).

We will take the equations $\left(\vphantom{\frac12}(\ref{mixsym})-({\ref{mixasym}})\right)$, (\ref{mixasym}), and (\ref{spatsym}) as our system,
\begin{equation}
\label{veceq}
\begin{array}{lc}
{\mathrm A}) & \bigtriangleup \left(\vphantom{\frac12}({\mathrm a}+b)\cdot {\mathcal M}_i + (b-{\mathrm a})\cdot \left({\mathcal L}_i +\dot\chi_i \right)\right) = 16 \pi G a(\tau) \cdot (\rho + p)\tilde{\mathfrak v}_i ,\\
{} & {}\\
{\mathrm B}) & (2c-{\mathrm a}-b)\cdot \left(\ddot{\mathcal L}_i + \ddot{\mathcal M}_i +2{\mathcal H} \left(\dot{\mathcal L}_i + \dot{\mathcal M}_i \right) - {\mathcal H}\bigtriangleup\chi_i\right) +\bigtriangleup \left(\vphantom{\dot{\mathcal L}_i}  2 ({\mathrm a}-b) \cdot {\mathcal L}_i +(2c-3b+{\mathrm a})\cdot\dot\chi_i \right)=0 ,\\
{} & {}\\
{\mathrm C})  & (2c-{\mathrm a}-b)\cdot \left(\dot{\mathcal L}_i + 3{\mathcal H}{\mathcal L}_i +\bigtriangleup \chi_i\right) + (2c+{\mathrm a}+b)\cdot \dot{\mathcal M}_i  + (6c+{\mathrm a}+b)\cdot {\mathcal H}{\mathcal M}_i=0,
\end{array}
\end{equation}
together with the law (\ref{veccons}) of the source term. Neglecting the equation (\ref{spatasym}) is a safe deed then, for it directly follows from other equations due to the Bianchi identity (\ref{vecBia}) and the matter conservation (\ref{veccons}).

In the equation (\ref{veceq}A), we immediately see a constraint which describes the influence of matter. Otherwise, generically all the rest is dynamical. Indeed, after removing a constrained variable, we can solve for the higher derivatives. By subtracting the equation (\ref{veceq}B) from the time derivative of equation (\ref{veceq}C) we find
\begin{equation}
\label{M}
\ddot{\mathcal M}_i +\frac{2c+3{\mathrm a}+3b}{2({\mathrm a}+b)}\cdot {\mathcal H} \dot{\mathcal M}_i +\frac{6c+{\mathrm a}+b}{2({\mathrm a}+b)}\cdot \dot{\mathcal H} {\mathcal M}_i=\frac{{\mathrm a}-b}{{\mathrm a}+b}\cdot \bigtriangleup\left({\mathcal L}_i +\dot\chi_i \right) + \frac{{\mathrm a}+b-2c}{2({\mathrm a}+b)}\cdot \left({\mathcal H}\left(\dot{\mathcal L}_i +\bigtriangleup\chi_i \right) + 3\dot{\mathcal H} {\mathcal L}_i\right).
\end{equation}
In the case of vacuum in Minkowski it used to imply $\ddot{\mathcal M}_i- \bigtriangleup {\mathcal M}_i =0$ by the virtue of (\ref{veceq}A). On the other hand, by time-differentiating the equation (\ref{veceq}A) now and using the conservation (\ref{veccons}), we derive
\begin{equation}
\label{L}
\dot{\mathcal L}_i=-\ddot\chi_i + \frac{{\mathrm a}+b}{{\mathrm a}-b}\cdot \left(\dot{\mathcal M}_i + 2{\mathcal H}{\mathcal M}_i\right) - 2{\mathcal H} \left({\mathcal L}_i +\dot\chi_i \right)
\end{equation}
which by substitution into equation (\ref{veceq}C) produces
\begin{equation}
\label{chi}
\ddot\chi_i +2{\mathcal H} \dot\chi_i - \bigtriangleup \chi_i= \frac{2(2{\mathrm a}c-b({\mathrm a}+b))}{({\mathrm a}-b)(2c-{\mathrm a}-b)}\cdot \dot{\mathcal M} +{\mathcal H} \left(  {\mathcal L}_i + \frac{2c(5{\mathrm a}-b)-({\mathrm a}^2+4{\mathrm a}b+3b^2)}{({\mathrm a}-b)(2c-{\mathrm a}-b)}\cdot {\mathcal M}_i \right).
\end{equation}
By using equations (\ref{veceq}A) and (\ref{L}), we can exclude the variable ${\mathcal L}_i$ and solve the remaining system (\ref{chi}, \ref{M}) for the accelerations $\ddot\chi_i$ and then $\ddot {\mathcal M}_i$. All in all, generically we have two dynamical vectors and one constrained vector, with no gauge symmetry left, precisely as we also had it around Minkowski space \cite{wedof}.

\subsubsection{Special cases}

Let us discuss all the cases of model parameter choices according to terminology of our previous paper \cite{wedof}. As is mentioned above, for the vectors we will give more details even for clearly unphysical cases, because now we can illustrate the methods in a much more compact way than it would be for the scalars.

{\bf Types 1 and 5.} Both types are in the general situation from above, for the restriction of $6c={\mathrm a}+b$  plays no role for the vectors. Other cases are not difficult to analyse either.

{\bf Type 2.} This is the case of  $2c={\mathrm a}+b$, or the so-called 1-parameter\footnote{It looks like having two parameters because we do not fix the value of effective gravitational constant.} New GR. Compared to GR, the constraint equation (\ref{veceq}A) doesn't undergo any change\footnote{thanks to equation (\ref{veceq}B)}, while the other equations (\ref{veceq}B, \ref{veceq}C) take the form of
$$\dot\chi_i + {\mathcal L}_i=0 \qquad \mathrm{and} \qquad \dot{\mathcal M}_i + 2{\mathcal H}{\mathcal M}_i =0.$$
Note their perfect correspondence with the covariant conservation (\ref{veccons}), or ${\mathcal M}_i \propto \frac{1}{a^2(\tau)}$. The numbers of the modes are the same as it was around Minkowski. 

The metric behaves absolutely in the GR way, while one half of the Lorentz sector is pure gauge with another half being constrained. In other words, the only difference from GR is in the less amount of gauge freedom in the Lorentzian modes, not directly observable though.

{\bf Types 3 and 7.} The case of ${\mathrm a}=b$ requires a bit more effort. The metric part is still absolutely the same as in GR (\ref{veceq}A):
$${\mathrm a} \bigtriangleup {\mathcal M}_i = 8 \pi G a(\tau) \cdot (\rho + p)\tilde{\mathfrak v}_i ,$$
hence 
$$\dot{\mathcal M}_i = - 2{\mathcal H} {\mathcal M}_i.$$ 
Substituting it into the other equations (\ref{veceq}B, \ref{veceq}C), we find
\begin{equation}
\label{3L1}
\ddot{\mathcal L}_i + 2{\mathcal H} \dot{\mathcal L}_i + \bigtriangleup \left(\dot\chi_i - {\mathcal H}\chi_i\right) - 2\dot{\mathcal H} {\mathcal M}_i=0,
\end{equation}
\begin{equation}
\label{3L2}
\dot{\mathcal L}_i + {\mathcal H}\left(3 {\mathcal L}_i + {\mathcal M}_i\right)  + \bigtriangleup \chi_i =0.
\end{equation}
Therefore, there is a constraint
\begin{equation}
\label{3chi}
\bigtriangleup \chi_i = - \dot{\mathcal L}_i - {\mathcal H} \left(3{\mathcal L}_i + {\mathcal M}_i \right).
\end{equation}

Up to now, everything is absolutely in line with the case of Minkowski. However, in that case an analogue of the constraint (\ref{3chi}) was the whole information, leaving one Lorentzian vector pure gauge. To the contrary, now we can substitute $\bigtriangleup \chi_i $ found from the second equation (\ref{3L2}) into the first one (\ref{3L1}):
\begin{equation}
\label{ohcon}
3\left({\mathcal H}^2-\dot{\mathcal H}\right) \left({\mathcal L}_i +  {\mathcal M}_i\right) =0.
\end{equation}
In Minkowski, as well as in de Sitter, it disappears leaving the extra gauge freedom alive. Otherwise, the whole sector is fully constrained. Indeed, with no need for initial data, we find ${\mathcal M}_i$ from (\ref{veceq}A), then ${\mathcal L}_i$ from (\ref{ohcon}), and finally $\chi_i$ from (\ref{3chi}). In other words, here we detect a kind of strong coupling\footnote{We use this term in a loose meaning, as any mismatch in numbers of pure gauge, constrained, or dynamical modes.} in Minkowski, in the form of accidental gauge symmetry. In the cosmological background, all three vectors are physical even though constrained, while in Minkowski, or in de Sitter, one of them was pure gauge.

{\bf Type 4.} The remaining codimension 1 case of the parameter space, ${\mathrm a}=-b$, is quite unphysical. However, the story is very similar to the previous one. The equation (\ref{veceq}A) constrains the ${\mathcal L}_i + \dot\chi_i$ combination, while the equation (\ref{veceq}C) gives yet another constraint:
\begin{equation}
\label{4ha}
\dot{\mathcal L}_i + \dot{\mathcal M}_i = - 3{\mathcal H} \left({\mathcal L}_i + {\mathcal M}_i\right) - \bigtriangleup\chi_i.
\end{equation}
In Minkowski, we could treat it as $\chi_i$ being gauge-free with $\mathcal L_i$ and $\dot{\mathcal M}_i$ given in terms of it. However, now we substitute the constraint (\ref{4ha}) into the formula (\ref{veceq}B) and, if not in Minkowski or de Sitter, get yet another constraint

\begin{equation}
\label{4oh}
3c\cdot \left({\mathcal H}^2-\dot{\mathcal H} \right) \left({\mathcal L}_i + {\mathcal M}_i\right) = 2b\cdot \bigtriangleup \left({\mathcal L}_i + \dot\chi_i\right) 
\end{equation}
which kills all the gauge freedom. The gravity variables are fully constrained. The equation (\ref{4oh}) together with the (\ref{veceq}A) one fix the quantity ${\mathcal L}_i + {\mathcal M}_i$ fully in terms of the matter content $(\rho +p)\tilde{\mathfrak v}_i$, and then other equations do so for $\chi_i$  and the vectors ${\mathcal L}_i $ and ${\mathcal M}_i $ separately.

{\bf Type 6.} The case of GR with ${\mathrm a}=b=c$ obviously brings us back to the standard
$$\bigtriangleup {\mathcal M}_i \propto a(\tau)\cdot  (\rho + p) \tilde{\mathfrak v}_i \propto \frac{1}{a^2(\tau)}$$
behaviour with the total gauge freedom of the Lorentzian sector.

{\bf Type 8.}  In the case of ${\mathrm a}=b=0$ the equation (\ref{veceq}A) immediately prohibits any vectorial ideal fluid perturbations. The equation (\ref{veceq}C) gives the result analogous to the one around Minkowski:
$$\bigtriangleup\chi_i = - \left(\dot{\mathcal L}_i + \dot{\mathcal M}_i \right) - 3{\mathcal H} \left({\mathcal L}_i + {\mathcal M}_i\right).$$
All the rest was pure gauge there. However, now we can substitute this result into the equation (\ref{veceq}B):
$$ \left({\mathcal H}^2-\dot{\mathcal H} \right) \left({\mathcal L}_i + {\mathcal M}_i\right) =0$$
which, away from Minkowski and de Sitter, kills half of remaining gauge freedom.

{\bf Type 9.} Finally, for the case of ${\mathrm a}=-b$ and $c=0$, the equations (\ref{veceq})
$$b \bigtriangleup \left({\mathcal L}_i + \dot\chi_i\right) = 8 \pi G a(\tau) \cdot (\rho + p)\tilde{\mathfrak v}_i =0$$
keep their internal behaviour absolutely the same as around Minkowski, however fully prohibit any vectorial ideal fluid perturbation.

\section{The scalar sector}

We come to the scalar sector, phenomenologically the most interesting one if to consider these models for the real world cosmology. In our gauge (\ref{ourgauge}), even the perturbed metric is diagonal
$$a^2(\tau)\cdot \left(\vphantom{\int}(1+2\phi)d\tau^2 - (1-2\psi)dx_i^2\right)$$ 
while the torsion components \cite{cosmo1} are
\begin{equation}
\label{scaltor}
\begin{array}{rcl}
T_{00i} & = & a^2 \cdot \partial_i \left({\dot\beta} - \phi +{\mathcal H} (\beta - \zeta)\right)\\
{} & {} & \xrightarrow{\text{gauge fixing}}  a^2 \cdot \partial_i \left({\dot\zeta} - \phi\right)\\
T_{ijk} & = &  a^2 \cdot \left(\delta_{ik}\partial_j \psi - \delta_{ij}\partial_k \psi + \epsilon_{ijl} \partial^2_{kl} s - \epsilon_{ikl} \partial^2_{jl} s \right)\\
T_{i0j} & = & -a^2 \cdot  {\mathcal H} \delta_{ij} +a^2 \cdot \left({\dot\psi} \delta_{ij} - \partial^2_{ij} {\dot\sigma}+ 2{\mathcal H}(\psi \delta_{ij}- \partial^2_{ij} \sigma) + \partial^2_{ij} \zeta  - \epsilon_{ijk}\partial_k {\dot s} \right) \\
{} & {} & \xrightarrow{\text{gauge fixing}} -a^2 \cdot {\mathcal H} \delta_{ij} +a^2 \cdot \left({\dot\psi} \delta_{ij} + 2{\mathcal H}\psi \delta_{ij} + \partial^2_{ij} \zeta - \epsilon_{ijk}\partial_k {\dot s} \right) 
\end{array}
\end{equation}
for the tensor and
\begin{equation}
\label{scalvec}
\begin{array}{rcl}
T_0 & = & 3{\mathcal H} + \bigtriangleup \left({\dot\sigma} - \zeta\right) -3 \dot\psi\\
{} & {} & \xrightarrow{\text{gauge fixing}} 3{\mathcal H} - \bigtriangleup\zeta -3 \dot\psi\\
T_i & = &  \partial_i \left(\phi - {\dot\beta} - 2\psi\right)\\
{} & {} & \xrightarrow{\text{gauge fixing}} \partial_i \left(\phi - {\dot\zeta} - 2\psi\right) 
\end{array}
\end{equation}
for the vector.

In order to have better control over what is going on, we again recall the covariant conservation of the energy-momentum of the matter content. In case of the ideal fluid (\ref{EMT}), its spatial components take the form of
$$\frac{1}{a^4(\tau)} \partial_0\left( a^4(\tau)\cdot (\rho + p)\frac{{\mathfrak v}_i}{a(\tau)} \right) = \partial_i \left(\vphantom{\frac12}\delta p + (\rho +p) \phi\right)$$
whose vectorial part (\ref{veccons}) has already been used in the previous section. In the scalar sector it means
\begin{equation}
\label{icons}
\frac{1}{a^4(\tau)} \partial_0\left( a^4(\tau)\cdot (\rho + p)\frac{{\mathfrak v}}{a(\tau)} \right) = \delta p + (\rho +p) \phi.
\end{equation}
At the same time, the time component of conservation is
\begin{equation}
\label{0cons}
\delta\dot\rho +3{\mathcal H} (\delta\rho + \delta p) = (\rho+p)\left(3\dot\psi + \frac{\bigtriangleup\mathfrak v}{a(\tau)}\right).
\end{equation}

\subsection{The weak gravity case}

By taking the partial-derivative divergence ${\mathfrak G}_{\mu\nu}\approx \partial_{\alpha} {\mathfrak S}_{\mu\nu}^{\hphantom{\mu\nu}\alpha}$ of the superpotential again, we reproduce the Minkowski space results \cite{wedof}
$${\mathfrak G}_{00}=\bigtriangleup \left(2c\cdot \psi + \frac{2c-{\mathrm a}-b}{2}\cdot \left({\dot\zeta}-\phi\right)\right),$$
$${\mathfrak G}_{0i}=\partial_i\left(2c\cdot \dot\psi + \frac{2c-{\mathrm a}-b}{2}\cdot \left({\ddot\zeta} - {\dot\phi}\right)\right),$$
$${\mathfrak G}_{i0}=\partial_i \left(\frac{6c-{\mathrm a}-b}{4}\cdot \dot\psi + \frac{2c-{\mathrm a}-b}{4}\cdot \bigtriangleup\zeta\right),$$
\begin{multline*}
{\mathfrak G}_{ij}=\left(\frac{6c -{\mathrm a}-b}{2}\cdot {\ddot\psi} +  \bigtriangleup \left(c\cdot\phi- \frac{4c -{\mathrm a}-b}{2}\cdot \psi\right)  \right)\delta_{ij} +\partial^2_{ij}\left(\frac{4c-{\mathrm a}-b}{2}\cdot \psi -c\cdot \phi +\frac{2c-{\mathrm a}-b}{2}\cdot {\dot\zeta}\right) \\
+ \frac{{\mathrm a}-b}{2}\cdot \epsilon_{ijk}\partial_k ({\ddot s} - \bigtriangleup s).
\end{multline*}
The appropriate redundancy of the corresponding equations of motion is immediately clear from the rather elementary identities of $\dot{\mathfrak G}_{00}\equiv\partial_i{\mathfrak G}_{0i}$ and $\dot{\mathfrak G}_{i0}\equiv\partial_j{\mathfrak G}_{ij}$. Restoration of GR in case of ${\mathrm a}=b=c$ is also obvious.

\subsection{Conformal transformation to cosmology}

The ${\mathfrak G}^0{}_0$ component transforms (\ref{congein}) in the same way as in GR, modulo the coefficient of $\frac{6c-{\mathrm a}-b}{2}$ instead of $2$, 
$$a^2(\tau )\cdot \underline{{\mathfrak G}^0{}_0} = \frac{6c-{\mathrm a}-b}{4}\cdot 3{\mathcal H}^2 + {\mathfrak G}^0{}_0 -\frac{3(6c-{\mathrm a}-b)}{2} \cdot \left({\mathcal H}\dot\psi + {\mathcal H}^2 \phi\right)$$
and produces the equation
\begin{equation}
\label{00}
 -3(6c-{\mathrm a}-b) \cdot \left({\mathcal H}\dot\psi + {\mathcal H}^2 \phi\right) + \bigtriangleup \left(4c\cdot \psi + (2c-{\mathrm a}-b)\cdot \left({\dot\zeta}-\phi\right)\right) = 16\pi G a^2(\tau) \cdot \delta\rho.
\end{equation}

In the mixed components, analogously to the vector sector, both symmetric and antisymmetric parts get the contribution of  $\frac{2c-{\mathrm a}-b}{4}\cdot{\mathcal H} (3T_{00i}+T_i) = \frac{2c-{\mathrm a}-b}{2}\cdot {\mathcal H} \partial_i \left(\dot\zeta -\phi-\psi\right)$ via ${\mathfrak G}_{0i}$. Multiplying them both by four, we get the antisymmetric equation
\begin{equation}
\label{as0i}
(2c-{\mathrm a}-b)\cdot  \left(\ddot\zeta  -\dot\phi -\dot\psi +2{\mathcal H}\left(\dot\zeta -\phi-\psi\right) - \bigtriangleup\zeta\right) =0
\end{equation}
and
\begin{equation}
\label{s0i}
(2c-{\mathrm a}-b)\cdot \left(\ddot\zeta + 2{\mathcal H}\left(\dot\zeta - \psi \right) + \bigtriangleup\zeta -\dot\phi\right) +(10c-{\mathrm a}-b)\cdot \dot\psi + 8c\cdot {\mathcal H}\phi = 32\pi G a(\tau) \cdot \left(\rho + p\right){\mathfrak v}
\end{equation}
for the symmetic part where we have solved equations for scalars  of the $\partial_i {\mathbb S}_1=\partial_i {\mathbb S}_2$ type simply as ${\mathbb S}_1={\mathbb S}_2$.

Finally, the novel contribution to the spatial components goes in terms of $T_{i0j}$ and $T_0 \delta_{ij}$. We find
$$a^2(\tau )\cdot \frac12 \left(\underline{{\mathfrak G}_{ij}} - \underline{{\mathfrak G}_{ji}}\right)=\frac{{\mathrm a}-b}{2} \cdot \epsilon_{ijk}\partial_k \left(\ddot s + 2{\mathcal H}\dot s -\bigtriangleup s\right)$$
and solve the equation as
\begin{equation}
\label{asij}
({\mathrm a}-b) \cdot \left(\ddot s + 2{\mathcal H}\dot s -\bigtriangleup s\right)=0.
\end{equation}

Analogously multiplying the results by two, we get the symmetric part of the $\partial^2_{ij}{\mathbb S}_1 + {\mathbb S}_2 \delta_{ij} = {\mathbb S}_3 \delta_{ij}$ form and solve it as ${\mathbb S}_1=0$ due to off-diagonal components of the equations:
\begin{equation}
\label{offs}
(2c-{\mathrm a}-b)\cdot \left(\dot\zeta + 3{\mathcal H}\zeta\right) + (4c-{\mathrm a}-b)\cdot\psi - 2c\cdot\phi =0,
\end{equation}
and therefore ${\mathbb S}_2  = {\mathbb S}_3$:
\begin{multline}
\label{delij}
(6c-{\mathrm a}-b)\cdot \left(\ddot\psi +{\mathcal H} \left(\dot\phi + 2\dot\psi\right) + (2\dot{\mathcal H} + {\mathcal H}^2)\phi \right)\\
 +\bigtriangleup \left(\vphantom{\frac12} 2c\cdot\phi - (4c-{\mathrm a}-b)\cdot \psi - (2c-{\mathrm a}-b)\cdot {\mathcal H}\zeta\right)= 16\pi G a^2(\tau)\cdot \delta p.
\end{multline}
The whole system of equations (\ref{00}, \ref{as0i}, \ref{s0i}, \ref{asij}, \ref{offs}, \ref{delij}) is here, with the pseudoscalar mode $s$ fully decoupled from all the rest.

\subsection{Analysis of equations}

Precisely as around Minkowski, the pseudoscalar variable $s$ is a diffeo-invariant dynamical mode (\ref{asij}) as long as ${\mathrm a}-b\neq 0$, and is pure gauge otherwise. In this sense, its behaviour is as robust as that of the standard graviton polarisations $h_{ij}$. Below we discuss the other three variables we have for the scalars: $\phi$, $\psi$, and $\zeta$. 

It is again quite easy to see that the system (\ref{00}, \ref{as0i}, \ref{s0i}, \ref{offs}, \ref{delij}) is not overdetermined despite being five equations for three variables. Indeed, we find the combination of the left hand sides
\begin{equation}
\frac12 (\partial_0 + 2{\mathcal H}) \left(\boxed{\ref{s0i}} - \boxed{\ref{as0i}}\right) - \left( \boxed{\ref{delij}}+\bigtriangleup \boxed{\ref{offs}}\right) \equiv (6c-{\mathrm a}-b)\cdot ({\mathcal H}^2-\dot{\mathcal H}) \phi
\end{equation}
which must be compared with the conservations law (\ref{icons}) and the background equation of motion 
$$(6c-{\mathrm a}-b)\cdot ({\mathcal H}^2-\dot{\mathcal H}) = 16\pi G a^2(\tau)\cdot (\rho+p).$$
In other words, we can neglect the equation (\ref{s0i}), as long as the energy-momentun tensor is conserved. It allows us to discuss the dynamics without worrying about the velocity potential $\mathfrak v$ which can be later restored by looking at this very same equation (\ref{s0i}). The motion of fluid is anyway dictated by the gradients of pressure.

At the same time, there is yet another identical equality
\begin{equation}
\partial_{0} \boxed{\ref{00}} + {\mathcal H} \left( \boxed{\ref{00}}+3 \boxed{\ref{delij}}+\bigtriangleup \boxed{\ref{offs}}\right)-\frac12 \bigtriangleup\left(\boxed{\ref{as0i}}+\boxed{\ref{s0i}}\right)   \equiv (6c-{\mathrm a}-b)\cdot 3 ({\mathcal H}^2-\dot{\mathcal H}) \dot\psi
\end{equation}
which is, in turn, to be compared with the conservation relation (\ref{0cons}). As is the usual practice in cosmology, we will use the temporal (\ref{00}) and the diagonal spatial (\ref{delij}) equations only in the combination of 
\begin{equation}
\label{newconeq}
\boxed{\ref{delij}}={\mathfrak c}_s^2\cdot \boxed{\ref{00}}
\end{equation}
with the sound speed ${\mathfrak c}_s^2=\frac{\partial p}{\partial\rho}$ in the ideal fluid (\ref{EMT}), which is enough as long as the energy conservation (\ref{0cons}) is ensured. This new equation (\ref{newconeq}) will not describe any purely gravitational mode but rather an acoustic wave in the fluid.

All in all, we take the system of the following three equations:
\begin{equation}
\label{scaleq}
\begin{array}{lc}
{\mathrm A}) &  (2c-{\mathrm a}-b)\cdot  \left(\ddot\zeta  -\dot\phi -\dot\psi +2{\mathcal H}\left(\dot\zeta -\phi-\psi\right) - \bigtriangleup\zeta\right) =0,\\
{} & {}\\
{\mathrm B}) & (2c-{\mathrm a}-b)\cdot \left(\dot\zeta + 3{\mathcal H}\zeta\right) + (4c-{\mathrm a}-b)\cdot\psi - 2c\cdot\phi =0 ,\\
{} & {}\\
{\mathrm C})  & (6c-{\mathrm a}-b)\cdot \left(\ddot\psi +{\mathcal H}\left(\dot\phi + (2+3{\mathfrak c}_s^2)\dot\psi\right) + \left(2\dot{\mathcal H} + (1+3{\mathfrak c}^2_s){\mathcal H}^2\right)\phi - {\mathfrak c}^2_s \bigtriangleup\psi\right)\hphantom{blah-blah}\\
{} & {}\\
{} & \hphantom{blah-blah-blah-blah-blah-blah } + (2c-{\mathrm a}-b)\cdot \bigtriangleup\left(\dot\zeta + 2{\mathcal H}\zeta - {\mathfrak c}^2_s \left(\dot\zeta- \phi - \psi\right) \right) =0
\end{array}
\end{equation}
where the lines (\ref{scaleq}A) and (\ref{scaleq}B) are the equations (\ref{as0i}) and (\ref{offs}) respectively, while the contribution (\ref{scaleq}C) is obtained by using the equation (\ref{offs}) inside the acoustic wave equation (\ref{newconeq}). Note that in the 1-parameter model, $2c-{\mathrm a}-b=0$, all the equations fully reproduce the case of GR. The Lorentzian mode is pure gauge, while $\phi=\psi$ and the wave of Newtonian potentials propagates at the speed of ${\mathfrak c}_s^2$.

The models of type 9, with ${\mathrm a} + b = c =0$, are absolutely trivial in this sector, and therefore no non-trivial energy-momentum tensor components are allowed. Otherwise, the equation (\ref{scaleq}B) is always a constraint. If $2c-{\mathrm a}-b\neq 0$ and $6c-{\mathrm a}-b\neq 0$, then the other two equations can be solved for the second (i.e. highest) time derivatives,  $\ddot\zeta$ and $\ddot\psi$. However, we need to check that the constraint can be substituted into them without spoiling this property. If $c\neq 0$, then one can solve (\ref{scaleq}B) for $\phi$, and see that the only problem appears in the equation (\ref{scaleq}A) since the only second time derivative comes there as $\frac{{\mathrm a} + b}{2c}\cdot \ddot\zeta$. Therefore, the case of ${\mathrm a} + b=0$ must be considered separately.

 The case of $c=0$ requires extra attention, but it doesn't produce anything new. In this case we can solve the constraint (\ref{scaleq}B) for $\psi$. Upon substitution, the equation (\ref{scaleq}A) features $2\ddot\zeta$ acceleration term while the remaining equation (\ref{scaleq}C) turns out to be third order in time derivatives of $\zeta$. However, a combination of $\partial_0 (\ref{scaleq}{\mathrm A}) + 2 (\ref{scaleq}{\mathrm C})$ yields a second-order equation, and allows then to solve the system for $\ddot\zeta$ and $\ddot\phi$.

For generic model parameters (${\mathrm a} + b$ not equal to zero or $2c$ or $6c$), we have two dynamical modes and one constrained mode in these equations. The Lorentzian mode is the new gravitational degree of freedom, also seen around Minkowski, while another dynamical mode represents the acoustic waves in the matter contents of the Universe. The limit of ${\mathcal H}=0$ is smooth.

\subsubsection{Special cases}

It would be somewhat cumbersome, to present the full dynamical analysis of all the equations (\ref{scaleq}) for the three scalars ($\phi,\psi,\zeta$) for all the cases in every detail. However, the interesting cases are already obvious, and the general illustration of the methods has been given in the vector sector. Here is just a brief list of the properties. Below we do not care about whether ${\mathrm a}=b$ or not. Remember though that it determines whether $s$ is pure gauge or dynamical.

{\bf Types 1 and 3.} The general case and the case of ${\mathrm a}=b$ are considered above. All the three variables are physical, i.e. not pure gauge, and we see one gravitational dynamical degree of freedom and one acoustic wave. It was the same around Minkowski.

{\bf Types 2 and 6.} Requiring only $2c={\mathrm a}+b$ is already enough for reducing the whole set of equations (\ref{scaleq}) to the case of GR, with the standard acoustic wave in the fluid and the Newtonian potentials it entails, and the Lorentzian variable $\zeta$ being gauge free.

{\bf Types 4 and 8.} If ${\mathrm a}+b=0$, we solve the equation (\ref{scaleq}B) as 
$$ \phi=\dot\zeta +3 {\mathcal H} \zeta + 2\psi.$$ 
Around Minkowski, the remaining two equations reduce to only one piece of information leaving an extra gauge freedom. In the case of cosmology it is different. The equation (\ref{scaleq}A) becomes a constraint for the velocity of $\psi + {\mathcal H} \zeta$, while the equation (\ref{scaleq}C) gets the only acceleration of $\psi + {\mathcal H} \zeta$. Therefore, the remaining system of equations can be reduced to two constraints. However, there is no gauge freedom any longer.

This case is definitely of no physical interest, so let us give only a very brief hint at how it goes. Having got rid of $\phi$, if we substitute the equation $\partial_0 (\ref{scaleq}{\mathrm A})$ for the acceleration $\partial^2_0 (\psi + {\mathcal H} \zeta) $ into the equation (\ref{scaleq}C), we generically get another constraint on the very same velocity $\partial_0 (\psi + {\mathcal H} \zeta)$, different from that of (\ref{scaleq}A) itself.  Then the two constraints fully fix one combination of $\psi$ and $\zeta$ leaving only one Cauchy datum free which looks like half degree of freedom.

{\bf Types 5 and 7.} If $6c = {\mathrm a} +b$, we get from equation (\ref{scaleq}B) that 
$$\phi =- \psi - 2(\dot\zeta + 3{\mathcal H}\zeta).$$
The equation (\ref{scaleq}A) takes a form of a wave equation for $\zeta$, 
$$\ddot\zeta + 4{\mathcal H}\dot\zeta + 2(\dot{\mathcal H} +2{\mathcal H}^2)\zeta - \frac13 \bigtriangleup\zeta =0$$
with no other source. However, the equation (\ref{scaleq}C) then says 
$$(1-3{\mathfrak c}^2_s) \left(\dot\zeta + 2{\mathcal H}\zeta\right)=0.$$ 
In case of ${\mathfrak c}^2_s=\frac13$, the Lorentzian sector does have a dynamical mode. However, generically the wave equation reduces to 
$$\bigtriangleup\zeta=0.$$ 
In this sense, the flat space vacuum results are fully reproduced.

The case is for sure unphysical. However, it exhibits a curious communication between the matter and the Lorentzian sector. The cosmology was purely vacuum, with a full freedom of the scale factor $a({\tau})$. The Lorentzian $\zeta$ mode around this strange background is possible only when a scale-invariant (traceless energy-momentum tensor, ${\mathfrak c}^2_s=\frac13$) fluid starts living on top of the vacuum solution. 

{\bf Type 9.} We take ${\mathrm a}+b=c=0$. It's trivial. Since all the left hand sides vanish identically, no first order matter fields are allowed.

\section{Discussion}

The pure-tetrad conformal transformations (\ref{contetr}) have appeared to be very useful for studying cosmological applications of the New General Relativity models. Derivation of the perturbation equations from the rather elementary weak gravity limit turned out to borrow many technical details from the very well-known procedure in the case of GR. At the same time, by analysing the equations we conveniently see that many versions of these models do exhibit strong coupling issues around Minkowski. Let us briefly  summarise
\begin{center}
{\bf the modifications of going from Minkowski vacuum to cosmology} 
\begin{tabular}{c|c|c} 
\hline
Type 1 & generic parameters & no change in degrees of freedom \\ 
\hline
Type 2 & ${\mathrm a}+b =2c $ & no change in degrees of freedom \\
\hline
Type 3 & ${\mathrm a}=b$ & the vectorial pure gauge becomes constrained \\
\hline
Type 4 & ${\mathrm a}+b =0$ & vectorial and scalar pure gauges become constrained  \\
\hline
Type 5 & ${\mathrm a}+b =6c $ & almost no change (see the details about scalars) \\
\hline
Type 6 & ${\mathrm a}=b=c$ & no change in degrees of freedom    \\
\hline
Type 7 & ${\mathrm a}=b=3c$ & mostly, the vectorial pure gauge becomes constrained \\
\hline
Type 8 & ${\mathrm a}=b=0$ & vectorial and scalar pure gauges become constrained \\
\hline
Type 9 & ${\mathrm a}+b =c=0$ & no change in degrees of freedom \\
\hline
\end{tabular}
\end{center}
Note though that many of these models, except the first three cases (types 1, 2, 3) and GR itself (type 6), do anyway have the metric not fully predictable, and therefore are not suitable for coupling the usual matter which might then be responsible for breaking (some of) the extra gauge freedom. However, this drawback is also present in one of other models (type 3, ${\mathrm a}=b$).

Away from GR, only the general modification (type 1) and the popular 1-parameter (${\mathrm a}+b=2c$, type 2) cases show admissible perturbative behaviour and no discontinuity of the ${\mathcal H} \rightarrow 0$ limit. The former theory has the maximal expected amount of dynamical modes when the four diffeomorphisms hit twice \cite{wedof}: on top of the four gauge freedoms also four physical variables are constrained, i.e. eight first-class constraints. As to the latter model, there are some arguments against its number of degrees of freedom being universally stable \cite{discon}. 

In our opinion, it would be interesting to study this question in more detail. Though in any case, we would say that the 1-parameter New GR model is not very suitable for attacking the cosmological problems. Compared to GR, it gets a new dynamical degree of freedom in the pseudo-scalar perturbation $s$ as well as a constraint $\dot\chi_i + {\mathcal L}_i=0$ in the vector part of the Lorentzian sector. However, the linear metric perturbations evolve independently and in absolutely the same way as in GR. Recalling also the situation with the static spherically symmetric solutions \cite{wespher, HaSh}, one might say it is a very boring modification of gravity which requires quite some work for deducing any observable deviation from GR\footnote{One possible way to go is to choose unnaturally complicated tetrad structures for the very simple spacetimes, though formally satisfying the symmetry requirements \cite{crtet}.}.

We consider the most general model, type 1 with no restriction on the parameters, as the most promising one for a viable modified gravity framework, even though it was previously refused \cite{HaSh, Ortin} precisely for not being boring enough. We still think that it would be good to have solutions different from GR. At the same time, there was yet another objection to it, in terms of ghosts \cite{Nwh}. Even though some researchers are now thinking of making friends with ghosts \cite{Vikman}, on our side we tend to take them as a serious problem. However, the analysis in the paper \cite{Nwh} does not seem reliable to us. 

We have explicitly calculated \cite{wedof} the kinetic matrix in the Newtonian gauge, and it can easily be made positive-definite (when ${\mathrm a}>|b|$ and $c>\frac{{\mathrm a}+b}{2}$), except for the conformal mode which is constrained and is anyway a ghost even in GR. Of course, it only means non-negativeness of the quadratic in momenta part of the Hamiltonian, while most probably there are also linear in momenta contributions, like for a usual vector field, which require more attention.

There are several issues with the approach taken by the paper \cite{Nwh}. On a milder side, it tries to avoid also the "tachyons" mistakenly attributing them to faster-than-light propagation. Being non-derivative negative energies, they do not necessarily mean any big trouble for predictability and/or cosmology. More importantly, the fight against the ghosts starts there from removing "nonlocal" models which means higher order poles in the propagator. How do they come in a model with a Lagrangian purely quadratic in velocities and with no higher derivatives? Those pathological poles are brought about by use of the spin formalism \cite{Nwh} which takes the projection operators and therefore adds more derivatives to the cake. That's why we disagree with the strict exclusion of such models from the list of potentially physical ones.

At the same time, even if the kinetic matrix did not have any negative eigenvalue at all, it wouldn't in itself imply guaranteed stability. For the Proca field, for instance, the momenta $\pi_i$ enter the canonical Hamiltonian as $\frac12 \pi^2_i + \pi^i \partial_i A_0$, independently of the sign of its mass-squared $m^2$. In this case, the spin formalism would basically mean doing the St{\"u}ckelberg trick by separating the longitudinal mode which gets endowed with the kinetic term $\frac12 m^2 {\dot\varphi}^2$. Equations of motion are equivalent before and after the trick, as long as we remember that $\varphi$ is not important in itself, but rather in its gradient $\partial_{\mu}\varphi$. A negative value of $m^2$ entails a full-fledged classical ghost in the representation of this model after the trick. Even for the Proca field, it is not quite clear whether this ghost is indeed as dangerous as it is usually assumed to be. And for the New GR analysis \cite{Nwh}, there is no proof that the spin formalism did not change the model. In any case, appearance of higher derivatives in the Lagrangian after a derivative change of variables does not automatically mean that the  model was bad.

We agree that the findings of the Ref. \cite{Nwh} are most probably rather problematic for quantising such models in the usual way. However, as a classical model of gravity, it might turn out  to be quite all right and calls for further careful investigation. Our conclusion is that the most general New GR models might be very promising for the modified gravity research, for they may finally provide us with an option of a modified teleparallel gravity without the ubiquitous strong coupling issues. If some modes do show bad instabilities, then one can think of other ways of adding constraints to the theory. Considering also the parity-odd terms could be a useful next step, too.

\end{document}